\documentclass[aps,prl,twocolumn,showpacs,superscriptaddress,groupedaddress]{revtex4}  
\usepackage{graphicx}  
\usepackage{dcolumn}   
\usepackage{bm}        
\usepackage{amssymb}   
\usepackage{slashed}
\usepackage{color}
\usepackage{lipsum}

\usepackage{amsmath,amsthm,amsfonts,amssymb,amscd}
\usepackage{physics}
\usepackage{multirow}
\usepackage{booktabs}
\usepackage{cancel}
\usepackage{color}
\usepackage{slashed}
\usepackage{subfigure}
\usepackage{dcolumn}
\usepackage{bm}

\begin{document}

\title{Possible large CP violation in charmed $\Lambda_b$ decays}

\author{Yin-Fa Shen\footnote{syf70280@hust.edu.cn}}
\address{School of Physics, Huazhong University of Science and Technology, Wuhan 430074, China}

\author{Jian-Peng Wang\footnote{wangjp20@lzu.edu.cn, corresponding author}}
\address{School of Nuclear Science and Technology,  Lanzhou University, Lanzhou 730000,  China}

\author{Qin Qin\footnote{qqin@hust.edu.cn,corresponding author}}
\address{School of Physics, Huazhong University of Science and Technology, Wuhan 430074, China}

\begin{abstract}
We propose that the cascade decay $\Lambda_b \to D(\to K^+\pi^-) N(\to p\pi^-)$ may serve as the discovery channel for baryonic CP violation. This decay chain is contributed by, dominantly, the amplitudes with the intermediate $D$ state as $D^0$ or $\bar{D}^0$. The large weak phase between the two kinds of amplitudes suggests the possibility of significant CP violation. While the presence of undetermined strong phases may complicate the dependence of CP asymmetry, our phenomenological analysis demonstrates that CP violation remains prominent across a broad range of strong phases. The mechanism also applies to similar decay modes such as $\Lambda_b \rightarrow D(\rightarrow K^+ K^-) \Lambda$. Considering the anticipated luminosity of LHCb, we conclude that these decay channels offer a promising opportunity to uncover CP violation in the baryon sector.
\end{abstract}

\maketitle

{\it Introduction.---}In modern particle physics, CP violation has gained increasing significance, particularly as one of the three Sakharov criteria for explaining the baryon asymmetry in the Universe~\cite{Sakharov:1967dj}. While the Cabibbo-Kobayashi-Maskawa (CKM) mechanism~\cite{Cabibbo:1963yz, Kobayashi:1973fv} in the standard model naturally incorporates CP violation, its magnitude is insufficient to fully account for the observed baryon asymmetry, which remains one of the most intriguing puzzles in contemporary physics~\cite{Bernreuther:2002uj, Canetti:2012zc}. It is worth noting that CP violation has been extensively studied in mesonic systems~\cite{Christenson:1964fg, NA31:1988eyf, BaBar:2001pki, Belle:2001zzw, Belle:2010xyn, BaBar:2010hvw, LHCb:2012xkq, LHCb:2013syl, LHCb:2019hro}, but has yet to be examined in any baryonic system. Therefore, given its closer connection to the baryon asymmetry in the Universe, the investigation of CP violation in the baryon sector remains a critical task in experimental research.

Many investigations have been conducted to search for CP violation in baryon decays, as evidenced by various studies~\cite{Belle:2021avh, Belle:2022uod, BESIII:2018cnd, BESIII:2021ypr, LHCb:2016yco, LHCb:2016rja, LHCb:2017slr, LHCb:2018fly, LHCb:2019jyj, LHCb:2019oke, Dai:2023zms}. These searches have introduced several intriguing and innovative techniques. For instance, the implementation of quantum entanglement in baryon-antibaryon pairs~\cite{BESIII:2018cnd, BESIII:2021ypr} provides opportunities to measure CP violation observables independent of strong phases. Additionally, nontraditional observables have been systematically developed and explored. Notably, the investigation of CP violation induced by time-reversal-odd correlations~\cite{Valencia:1988it, Datta:2003mj, Datta:2011qz, Gronau:2011cf, Gronau:2015gha, Wang:2022fih} provides well-defined and complementary observables for experimental CP-violation searches.

While CP violation in the baryon sector still requires further exploration in both the experimental and theoretical domains, there have been encouraging pieces of evidence of its presence in $\Lambda_b$ baryon decays~\cite{LHCb:2016yco}. Moreover, the LHCb experiment has not only successfully generated and collected unprecedented data on $b \bar b$ pairs but also has plans to increase its luminosity in the near future~\cite{LHCb:2008vvz, LHCb:2014set, LHCb:2018roe, LHCb:2023hlw}. Therefore, there is considerable optimism for the eventual discovery of CP violation in $b$ baryons at the LHCb.

In this study, we propose a potential decay channel for the discovery of baryonic CP violation: $\Lambda_b \to D(\to K^+\pi^-) N(\to p\pi^-)$, where $D$ represents a superposition of $D^0$ and $\bar{D}^0$, and $N$ can correspond to any excited state of the neutron. Specifically, we focus on $N(1440)$ and $N(1520)$ as examples in this investigation, which possess spin-1/2 and -3/2, respectively. Since the CP violation in this decay arises from the interference between $D^0$- and $\bar{D}^0$-contributing amplitudes with a significant weak phase difference, we anticipate a substantial level of CP violation. Furthermore, based on the previous measurements by LHCb~\cite{LHCb:2013jih, LHCb:2021ohr}, we estimate that approximately $\mathcal{O}(10^2)$ events can be collected after the Run 1-2 phase of LHCb, with even more expected during Run 3-4. This provides a promising opportunity for the experimental discovery of baryonic CP violation.

In the subsequent sections of this paper, we will begin by deriving the expressions for the CP asymmetry in the decay chain $\Lambda_b \to D(\to K^+\pi^-) N(\to p\pi^-)$. These expressions, formulated within the framework of helicity amplitudes~\cite{Jacob:1959at}, will illustrate the dependence of CP asymmetry on various strong phases. Subsequently, we will conduct phenomenological analyses to demonstrate that CP violation remains significant over a wide range of strong phases. Based on these compelling findings, we strongly recommend that LHCb investigates the CP violation in this specific decay channel.

\begin{figure}
    \includegraphics[keepaspectratio, width=7cm]{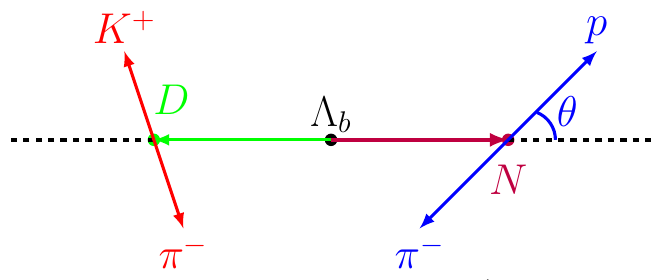}
    \caption{Sketch of the full decay chain $\Lambda_b \to D(\to K^+\pi^-) N(\to p\pi^-)$.}\label{fig:theta}
\end{figure}

{\it Theoretical setup.---}In this section, we investigate the CP violation of the decay chain $\Lambda_b \to D(\to K^+\pi^-) N(\to p\pi^-)$, as depicted in FIG.~\ref{fig:theta}. It is worth noting that the intermediate $D$ meson can manifest as either $D^0$ or $\bar{D}^0$, and the baryon $N$ donates excited states of neutrons contributing in the channel depending on the specific $p\pi^-$ invariant mass. For a particular $N$, two decay paths contribute to the desired final state, namely, $\Lambda_b \to \bar D^0(\to K^+\pi^-) N$ and $\Lambda_b \to D^0(\to K^+\pi^-) N$. The mixing effect of the neutral $D$ meson is neglected in our analysis due to its high suppression~\cite{ParticleDataGroup:2022pth, HFLAV:2022pwe}.

The relation between the decay amplitudes of $\Lambda_b$ and $D$ can be parametrized as $\braket{K^+\pi^-}{D^0} $ $ =  -r_D e^{-i\delta_D}\braket{K^+\pi^-}{\bar D^0}$ and $\braket{\bar D^0 N}{\Lambda_b} = r_B e^{i(\delta_B + \omega)} $ $\braket{D^0 N}{\Lambda_b}$, respectively. The parameters $r_D$ and $r_B$ are the relevant magnitude ratios, $\delta_D$ and $\delta_B$ are the relevant strong phases, $\omega$ is the weak phase between the $\Lambda_b$ decays, and the weak phase between the $D$ decays is neglected. As multiple $N$ states and various partial-wave amplitudes will contribute to $\Lambda_b\to DN$ decays, a series of $r_B$'s and $\delta_B$'s will be introduced in the subsequent calculations, as listed in TABLE~\ref{table:parameters}. Notably, both $r_B$ and $r_D$ are of the order $\mathcal{O}(10^{-2})$,  signifying that the two decay paths have comparable amplitude magnitudes and therefore result in considerable interference effects. In addition, the weak phase is substantial with $\sin{\omega}\approx$ 0.9, suggesting the potential for a significant CP-violation effect.

In practical analyses, the consideration of $N$-baryon resonances becomes inevitable due to the presence of the $p\pi^-$ final state. Furthermore, these resonance peaks are closely spaced, making them almost indistinguishable in experiments. To address this point, we take the $N$ baryon as a superposition of $N(1440)$ and $N(1520)$ for an illustrative example to investigate the complete decay chain $\Lambda_b \to D(\to K^+\pi^-) N(\to p\pi^-)$. For the spin-1/2 $N(1440)$ resonance, the angular momentum conservation allows $\Lambda_b \to D N(1440)$ to occur through both the $S$ wave and $P$ wave, characterized by amplitudes denoted as $S_{1/2}$ and $P_{1/2}$ in the subsequent analysis. In the case of the spin-3/2 $N(1520)$ resonance, the $\Lambda_b \to D N(1520)$ process can occur through both the $P$ wave and $D$ wave, with amplitudes represented as $P_{3/2}$ and $D_{3/2}$. Since the strong interaction conserves both parity (P) and charge-parity (CP), each secondary $N\to p\pi^-$ decay amplitude can be encoded by one single amplitude. Therefore, they can be directly absorbed into the aforementioned $S, P, D$ amplitudes, inducing no new parameters. 

Based on the formulas of the helicity framework~\cite{Jacob:1959at}, we can derive the differential decay width of ${\dd \Gamma}/{\dd\cos{\theta}}$ in terms of the partial-wave amplitudes, where $\theta$ is the angle between the proton motion direction in the $N$ rest frame and that of the $N$ in the $\Lambda_b$ rest frame, as illustrated in FIG.~\ref{fig:theta}. With also the differential decay width ${\dd \bar{\Gamma}}/{\dd\cos{\theta}}$ of the charge conjugate process, we define the $\theta$-dependent CP asymmetry as
\begin{equation}\label{ACP}
    A_{\rm{CP}} \equiv  { {\dd \Gamma}/{\dd\cos{\theta}}-{\dd \bar \Gamma}/{\dd\cos{\theta}} \over {\dd \Gamma}/{\dd\cos{\theta}}+ {\dd \bar \Gamma}/{\dd\cos{\theta}} } \equiv \frac{\mathcal{N}(\theta)}{\mathcal{D}(\theta)} \; ,
\end{equation}
where different polarization states in the final state are summed and in the initial state are averaged.

The numerator, denoted as $\mathcal{N}(\theta)$, reads
\begin{align}\label{Num}
        \mathcal{N}(\theta) =& 
        \left|S_{1/2}\right|^2 \mathcal{N}_{d} (r_{B1},\delta_{B1}) +\left|P_{1/2}\right|^2 \mathcal{N}_{d} (r_{B3},\delta_{B3}) \\
        &+ (1+3\cos^2{\theta}) \left|P_{3/2}\right|^2 \mathcal{N}_{d} (r_{B2},\delta_{B2}) \nonumber \\
        & + (1+3\cos^2{\theta}) \left|D_{3/2}\right|^2 \mathcal{N}_{d} (r_{B4},\delta_{B4})\nonumber\\
        &+ 2 \cos{\theta} \left|S_{1/2} P_{3/2}\right| \mathcal{N}_{i} (r_{B1}, r_{B2}, \delta_{B1}, \delta_{B2}, \delta_{PS})\nonumber\\
        &+ 2\cos{\theta} \left|P_{1/2} D_{3/2}\right| \mathcal{N}_{i} (r_{B3}, r_{B4}, \delta_{B3}, \delta_{B4}, \delta_{DP}),\nonumber
    \end{align}
where the $\mathcal{N}_{d}$ and $\mathcal{N}_{i}$ functions are defined by 
\begin{align}
    \mathcal{N}_{d}(r_{i},\delta_{i})=&\ 4 r_{i} r_D \sin{\omega}\sin{(\delta_D+\delta_{i})}, \\
    \mathcal{N}_{i}(r_1, r_2 ,\delta_{1},\delta_{2},\delta_{3})= &\ 4 r_D \sin{\omega} [ r_{1} \sin{(\delta_D + \delta_{1} - \delta_{3})}  \nonumber \\
 &   + r_{2} \sin{(\delta_D + \delta_{2} + \delta_{3})}. \nonumber
\end{align}
The newly introduced ratios $r_{Bi}$ and strong phases $\delta_{Bi}$ parametrize the relations between different partial-wave contributions to the $\Lambda_b$ decay amplitudes $\braket{\bar D^0 N}{\Lambda_b}_i = r_{Bi} e^{i(\delta_{Bi} + \omega)} \braket{D^0 N}{\Lambda_b}_i$, with $i=1,2,3,4$ corresponding to the $S_{1/2}$, $P_{3/2}$, $P_{1/2}$ and $D_{3/2}$, respectively. 
The denominator, denoted as $\mathcal{D}(\theta)$, is defined as
\begin{align}\label{Den}
        \mathcal{D}(\theta) =& 
        \left|S_{1/2}\right|^2 \mathcal{D}_{d} (r_{B1},\delta_{B1})  +\left|P_{1/2}\right|^2 \mathcal{D}_{d} (r_{B3},\delta_{B3}) \\
        &+ (3\cos^2{\theta}+1) \left|P_{3/2}\right|^2 \mathcal{D}_{d} (r_{B2},\delta_{B2}) \nonumber \\
        & + (3\cos^2{\theta}+1) \left|D_{3/2}\right|^2 \mathcal{D}_{d} (r_{B4},\delta_{B4})\nonumber \\
        &+ 2 \cos{\theta} \left|S_{1/2}P_{3/2}\right| \mathcal{D}_{i} (r_{B1}, r_{B2}, \delta_{B1}, \delta_{B2}, \delta_{PS})\nonumber \\
        &+ 2 \cos{\theta} \left|P_{1/2}D_{3/2}\right| \mathcal{D}_{i} (r_{B3}, r_{B4}, \delta_{B3}, \delta_{B4}, \delta_{DP}), \nonumber 
    \end{align}
where the $\mathcal{D}_{d}$ and $\mathcal{D}_{i}$ functions are defined by 
\begin{align}
  & \mathcal{D}_{d}(r_{i},\delta_{i}) = 2r^2_{i} + 2r^2_D 
    - 4 r_{i} r_D \cos{\omega}\cos{(\delta_{i}+\delta_D)}, \\
  & \mathcal{D}_{i}(r_1, r_2, \delta_{1}, \delta_{2}, \delta_{3}) =4 r_{1} r_{2} \cos{(\delta_{1}-\delta_{2}-\delta_{3})}  \nonumber \\
    & \qquad \qquad + 4r^2_D \cos{\delta_{3}} - 4  r_D \cos{\omega} [r_1 \cos{(\delta_D + \delta_{1} - \delta_{3})} \nonumber \\
    &\qquad \qquad  + r_2 \cos{(\delta_D + \delta_{2} + \delta_{3})}]. \nonumber 
    \end{align}

With unpolarized initial states, interferences between different partial-wave amplitudes of the same $N$ baryon vanish, eliminating terms proportional to ${\rm Re} [S_{1/2} P^\star_{1/2}]$ and ${\rm Re} [P_{3/2} D^\star_{3/2}]$, which are analogous to the Lee-Yang parameter $\alpha$~\cite{Lee:1957qs}. Consequently, only interferences between partial waves associated with different $N$ baryons survive. To formulate these, we introduce $\delta_{PS}$ as the strong phase between the $P_{3/2}$ and $S_{1/2}$ amplitudes, and $\delta_{DP}$ as the strong phase between the $D_{3/2}$ and $P_{1/2}$ amplitudes, namely, $\delta_{PS} = \arg [\braket{D^0 N}{\Lambda_b}_2/ \braket{D^0 N}{\Lambda_b}_1]$ and $\delta_{DP} = \arg [\braket{D^0 N}{\Lambda_b}_4/ \braket{D^0 N}{\Lambda_b}_3]$. It is evident that $\mathcal{N}_{d}$'s indicate the CP asymmetries of each independent partial wave, while $\mathcal{N}_{i}$'s reflect the CP-violation effects induced by interferences between different partial waves.

\begin{table}
\centering
\caption{The input parameters and their values.}\label{table:parameters}
{
\begin{tabular}{cc}
\hline
 \textbf{Parameter}    &  \textbf{Value} \\
\hline
  $r_0\equiv |V_{ub}V^\star_{cd}/V_{cb} V^\star_{ud}|$   &  $(2.0 \pm 0.1)\times 10^{-2}$~\cite{ParticleDataGroup:2022pth}\\
  $\omega = \arg{(V_{ub} V^\star_{cd}/V_{cb} V^\star_{ud})}$    &  $(114.4\pm 1.5)^\circ$~\cite{ParticleDataGroup:2022pth}\\
  $r_D$       &  $(5.86\pm 0.02)\times 10^{-2}$~\cite{ParticleDataGroup:2022pth}\\
  $\delta_D$   &    $\left(7.2^{+7.9}_{-9.2}\right)^\circ$~\cite{HFLAV:2022pwe}\\
\hline
\end{tabular}
}
\end{table}

{\it Phenomenology.---}
In the phenomenological analysis, we introduce two observables of interest. The first observable, denoted as $A_1$, is defined as 
\begin{equation}
    A_1 \equiv \frac{\int^{1}_{-1} \mathcal{N}(\theta) \; \dd \cos{\theta}}{\int^{1}_{-1} \mathcal{D}(\theta) \; \dd \cos{\theta}} \equiv \frac{\mathcal{N}_1}{\mathcal{D}_1},
\end{equation}
where the numerator $\mathcal{N}_1$ and denominator $\mathcal{D}_1$ are calculated to be 
\begin{align}
    \mathcal{N}_1 = &\ \abs{S_{1/2}}^2 \mathcal{N}_{d} (r_{B1},\delta_{B1}) + \abs{P_{1/2}}^2 \mathcal{N}_{d} (r_{B3},\delta_{B3})   \\
    &+ 2\abs{P_{3/2}}^2 \mathcal{N}_{d} (r_{B2},\delta_{B2}) + 2\abs{D_{3/2}}^2\mathcal{N}_{d} (r_{B4},\delta_{B4}) , \nonumber  \\
    \mathcal{D}_1 = &\  \abs{S_{1/2}}^2 \mathcal{D}_{d} (r_{B1},\delta_{B1}) + \abs{P_{1/2}}^2 \mathcal{D}_{d} (r_{B3},\delta_{B3}) \nonumber \\
    &+ 2 \abs{P_{3/2}}^2 \mathcal{D}_{d} (r_{B2},\delta_{B2}) + 2 \abs{D_{3/2}}^2\mathcal{D}_{d} (r_{B4},\delta_{B4}). \nonumber 
    \end{align}
It can be noticed that $A_1$ is expressed as the sum of contributions from four independent partial waves, regardless of the strong phases $\delta_{PS}$ and $\delta_{DP}$. This suggests that any interferences between different partial waves are entirely eliminated. The second observable, denoted as $A_2$, is designed to retain the interference of distinct partial waves. It is defined by 
\begin{equation}
    A_2 \equiv \frac{\int^{1}_{-1} \mathrm{sgn}[\cos{\theta}] \cdot \mathcal{N}(\theta) \; \dd \cos{\theta}}{\int^{1}_{-1} \mathcal{D}(\theta) \; \dd \cos{\theta}} \equiv \frac{\mathcal{N}_2}{\mathcal{D}_1},
\end{equation}
analogous to the CP violation induced by the forward-backward asymmetry~\cite{Zhang:2021zhr, Hu:2022eql, Zhang:2022emj, Wei:2022zuf, Zhang:2022iye}. Here, $\mathrm{sgn}[x]$ represents the sign function, and $\mathcal{N}_2$ is given by
\begin{align}
    \mathcal{N}_2 = &\ \left|S_{1/2}\right| \left|P_{3/2}\right|\mathcal{N}_{i} (r_{B1}, r_{B2},\delta_{B1},\delta_{B2},\delta_{PS}) \\
    &+ \left|P_{1/2}\right| \left|D_{3/2}\right|\mathcal{N}_{i} (r_{B3}, r_{B4},\delta_{B3},\delta_{B4},\delta_{DP}) . \nonumber 
    \end{align}

It can be observed from formula \eqref{Num} that six unconstrained strong phases and various ratios of amplitudes are involved in the prediction for the CP asymmetries. Consequently, conducting a comprehensive analysis presents a formidable challenge. To streamline our investigation, we leave these unknown phases and ratios as free parameters and show in most areas of the parameter space large CP violation is expected. In practice, we consider three cases, in which certain relations between the ratios and phases are assumed. We initially consider case 1 with a single resonance of the $N$ baryon and allow for the simultaneous contributions from $N(1440)$ and $N(1520)$ in case 2 and case 3. More detailed setups are described below.

\begin{widetext}

\begin{figure}[htbp]
    \centering
    \includegraphics[keepaspectratio,width=5.4cm]{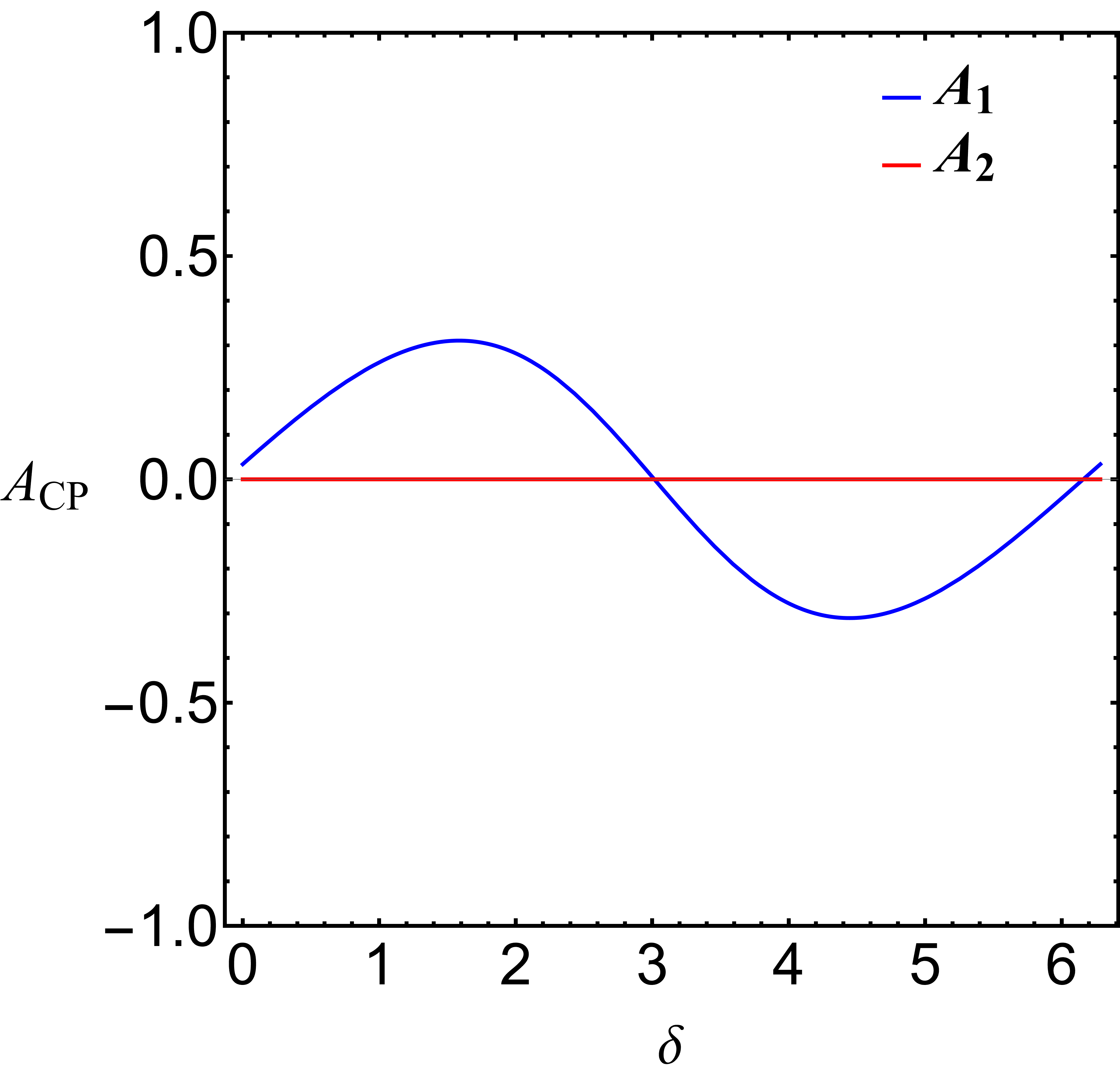}
    \includegraphics[keepaspectratio,width=5.4cm]{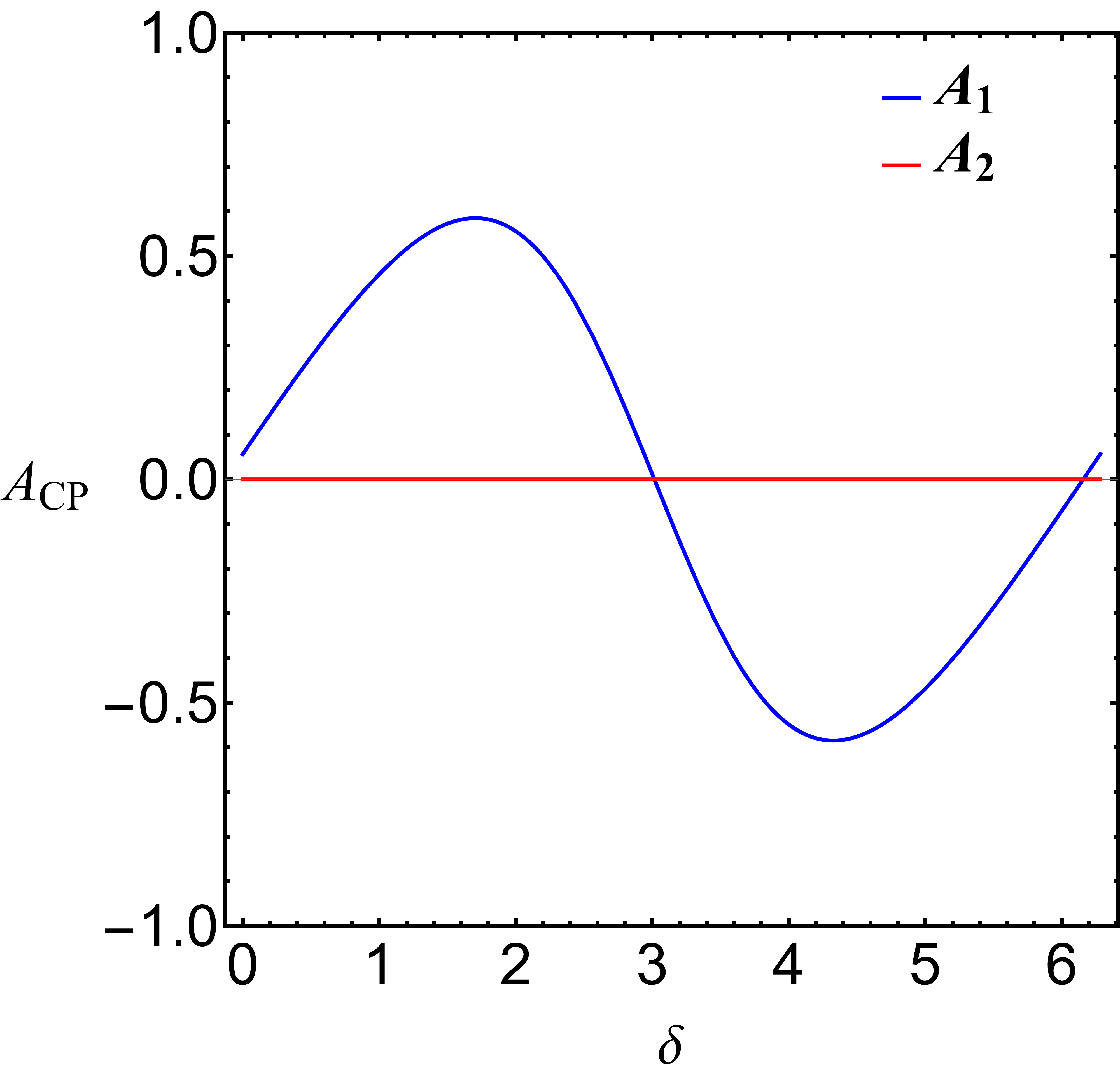}
    \includegraphics[keepaspectratio,width=5.4cm]{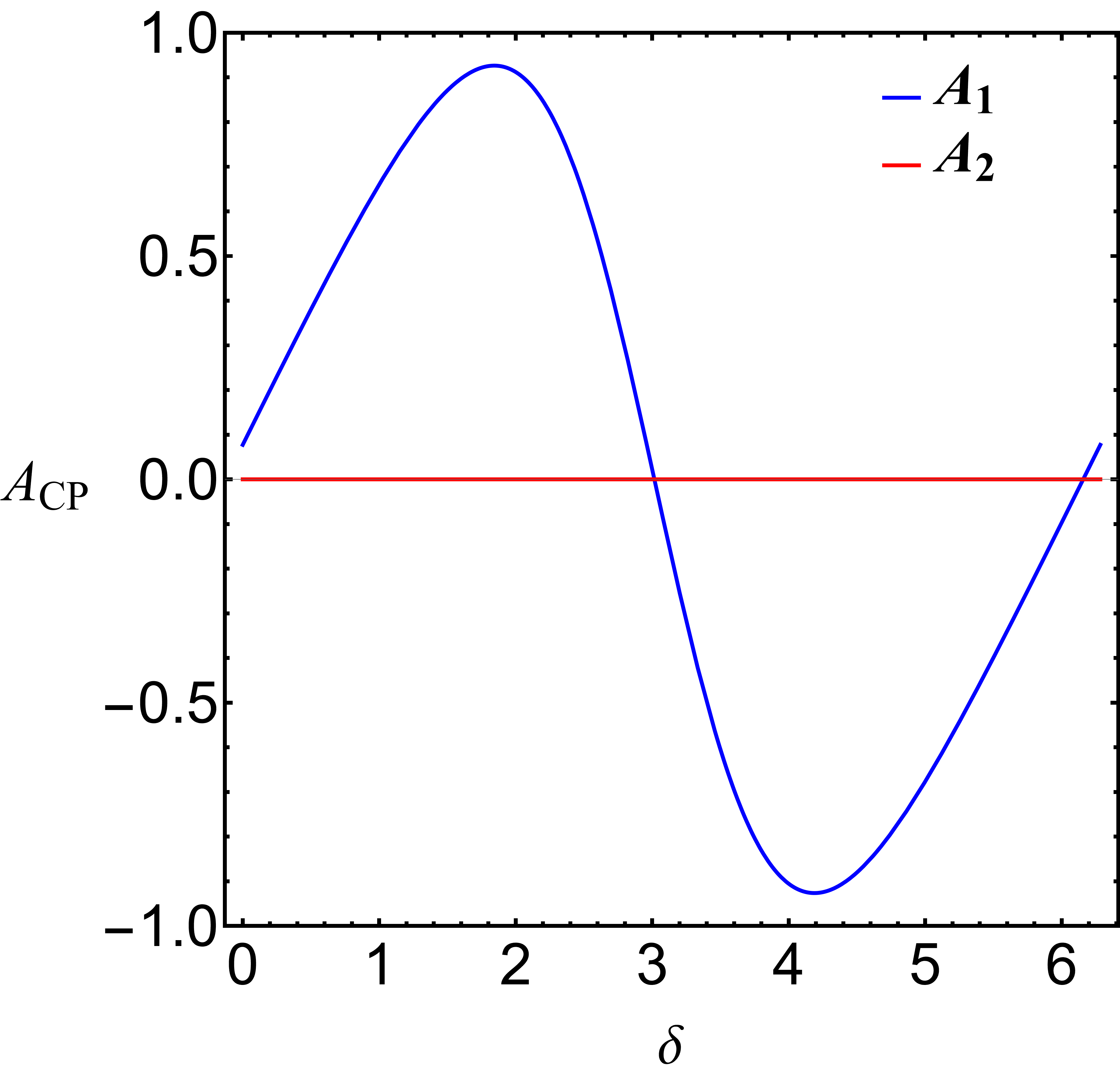}
    \caption{The dependence of the asymmetries $A_1$ and $A_2$ on the strong phase $\delta=\delta_{B1}=\delta_{B3}$ in case 1. The left, middle, and right panels correspond to the three benchmarks (a) $r_{B1}=r_{B3}=r_0/2$, (b) $r_{B1}=r_{B3}=r_0$, and (c) $r_{B1}=r_{B3}=2 r_0$.}\label{fig:Case1}
\end{figure}

In case 1, we initiate our investigation by examining a single $N$-baryon resonance, specifically when $|S_{1/2}|=|P_{1/2}|$ and $|P_{3/2}|=|D_{3/2}|=0$. Under these conditions, the asymmetry $A_2$ remains precisely zero. In addition, we set both involved strong phases to a common value, denoted as $\delta_{B1}=\delta_{B3}=\delta$. For the amplitude ratios, we consider three distinct benchmarks: (a) $r_{B1}=r_{B3}=r_0/2$, (b) $r_{B1}=r_{B3}=r_0$, and (c) $r_{B1}=r_{B3}=2 r_0$. The results for the  CP asymmetry $A_{1,2}$ as functions of $\delta$ are depicted in FIG.~\ref{fig:Case1}.

Across all three benchmarks, the CP asymmetry $A_1$ generally reaches magnitudes of $\mathcal{O}(10\%)$, with the exception of narrow regions of the strong phases. Notably, in the third benchmark, the peak value of $A_1$ can reach an impressive magnitude of approximately 90\%. 

\begin{figure}[htbp]
    \centering
    \includegraphics[keepaspectratio,width=5.4cm]{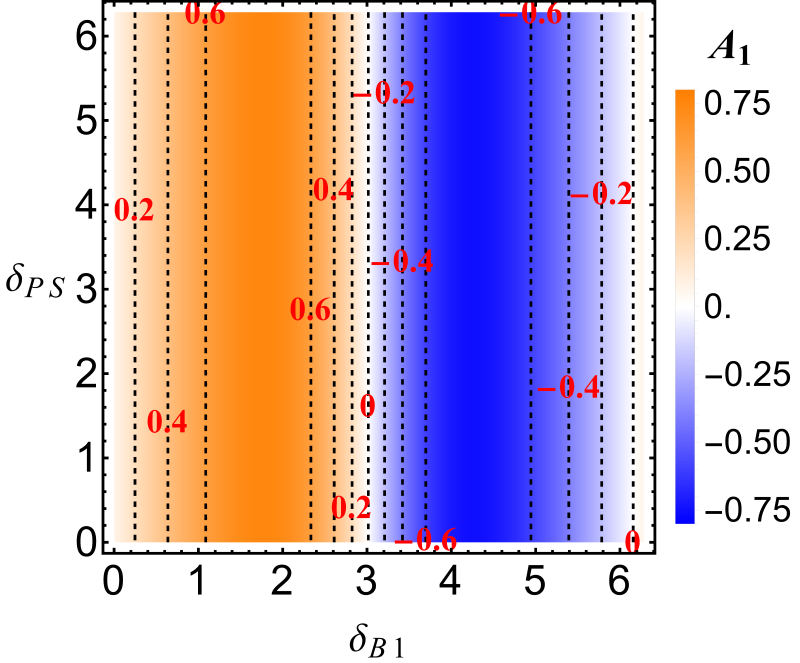}
    \includegraphics[keepaspectratio,width=5.4cm]{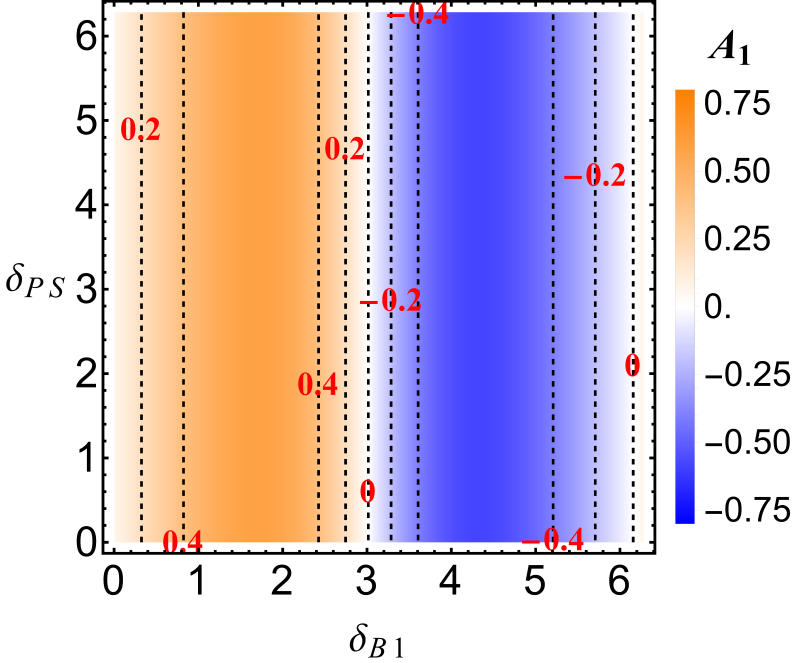}
    \includegraphics[keepaspectratio,width=5.4cm]{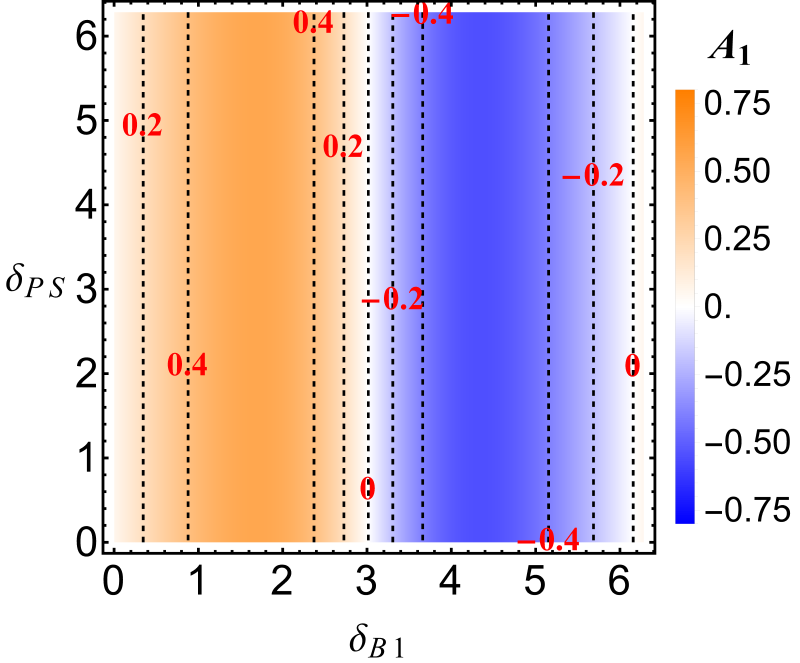}
    \caption{Two-dimensional dependence of the asymmetry $A_1$ on the strong phases in case 2. The left, middle, and right panels correspond to the three benchmarks (a) $r_{B1}=r_{B3}= r_0/2, r_{B2}=r_{B4}=2 r_0$, (b) $r_{B1}=r_{B2}=r_{B3}=r_{B4}=r_0$, and (c) $r_{B1}=r_{B3}=2 r_0, r_{B2}=r_{B4}= r_0/2$. The dashed lines on the graph represent contours with associated red values of $A_1$.}\label{fig:Case2A1}
\end{figure}

\begin{figure}
    \centering
    \includegraphics[keepaspectratio,width=5.4cm]{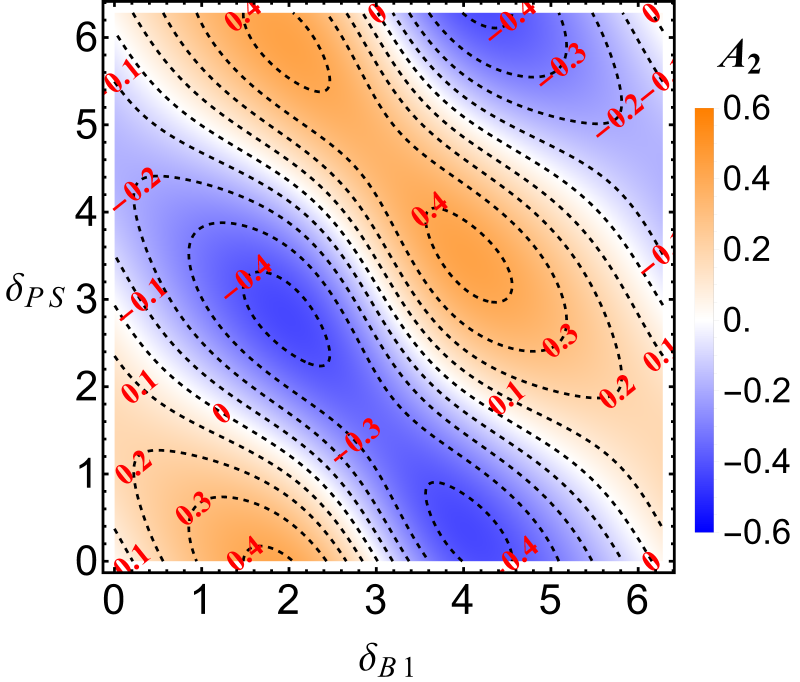}
    \includegraphics[keepaspectratio,width=5.4cm]{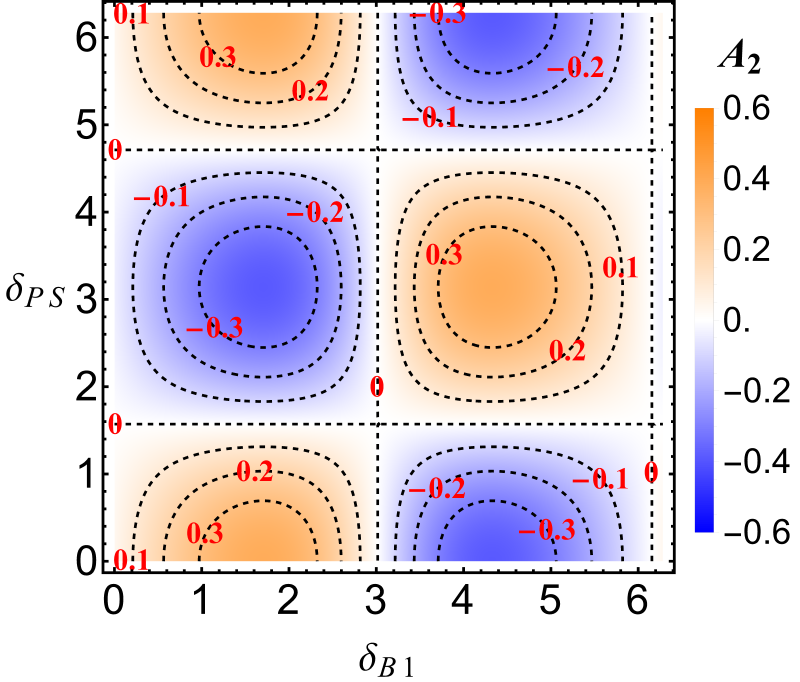}
    \includegraphics[keepaspectratio,width=5.4cm]{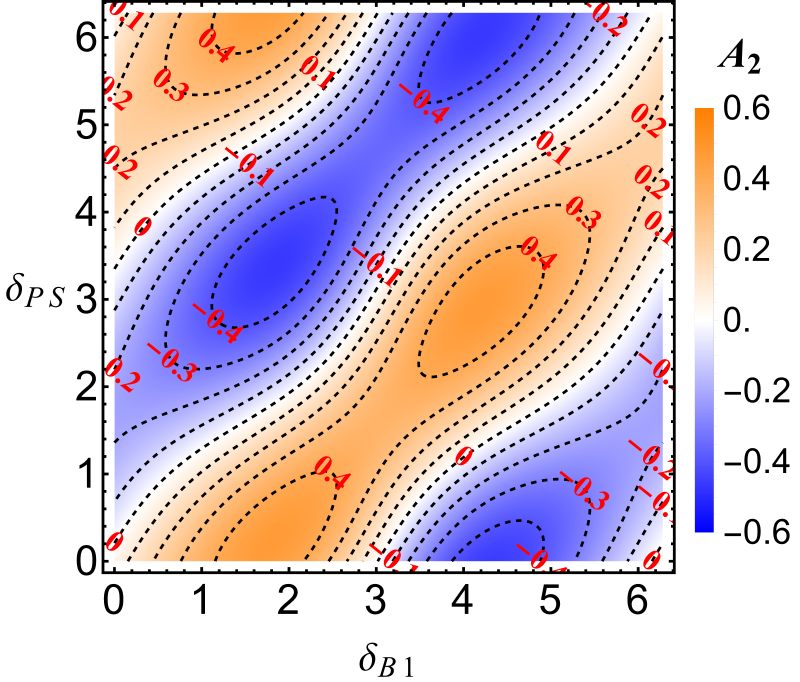}
    \caption{Two-dimensional dependence of the asymmetry $A_2$ on the strong phases in case 2. The left, middle, and right panels correspond to the three benchmarks (a) $r_{B1}=r_{B3}= r_0/2, r_{B2}=r_{B4}=2 r_0$, (b) $r_{B1}=r_{B2}=r_{B3}=r_{B4}=r_0$, and (c) $r_{B1}=r_{B3}=2 r_0, r_{B2}=r_{B4}= r_0/2$. The dashed lines on the graph represent contours with associated red values of $A_2$.}\label{fig:Case2A2}
\end{figure}

In case 2, subsequently, we consider both $N(1440)$ and $N(1520)$ contributing with $|S_{1/2}|=|P_{1/2}|=|P_{3/2}|=|D_{3/2}|$. Furthermore, we suppose that $\delta_{B1}=\delta_{B2}=\delta_{B3}=\delta_{B4}$ and $\delta_{PS}=\delta_{DP}$, allowing for two independent strong phases, and select three benchmarks for $\{r_{Bi}\}$: (a) $r_{B1}=r_{B3}= r_0/2, r_{B2}=r_{B4}=2 r_0$, (b) $r_{B1}=r_{B2}=r_{B3}=r_{B4}=r_0$, and (c) $r_{B1}=r_{B3}=2 r_0, r_{B2}=r_{B4}= r_0/2$. The dependencies of $A_{1,2}$ on the two strong phases are displayed in FIG.~\ref{fig:Case2A1} and FIG.~\ref{fig:Case2A2}, respectively. Notably, $A_1$ remains completely unaffected by the strong phases between different partial waves, yielding results similar to those presented in FIG.~\ref{fig:Case1}.

The behavior of $A_1$ closely resembles that observed in case 1. Because of the interferences between the $N(1440)$ and $N(1520)$ amplitudes, $A_2$ is also nonzero and significant across a majority of the parameter space defined by the strong phases. Therefore, $A_2$ serves as an additional observable to complement $A_1$, particularly in situations where $A_1$ is small. 

\begin{figure}[htbp]
    \centering
    \includegraphics[keepaspectratio,width=5.4cm]{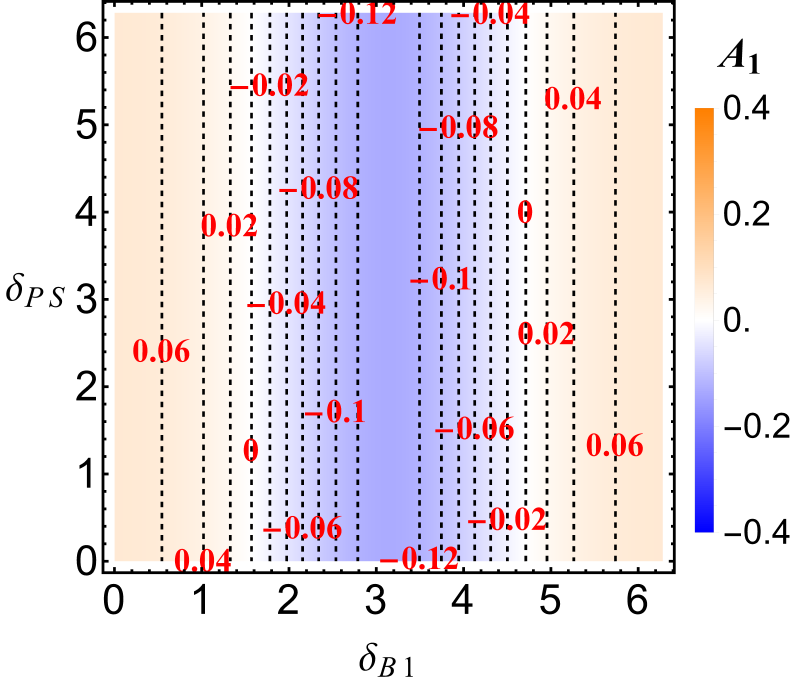}
    \includegraphics[keepaspectratio,width=5.4cm]{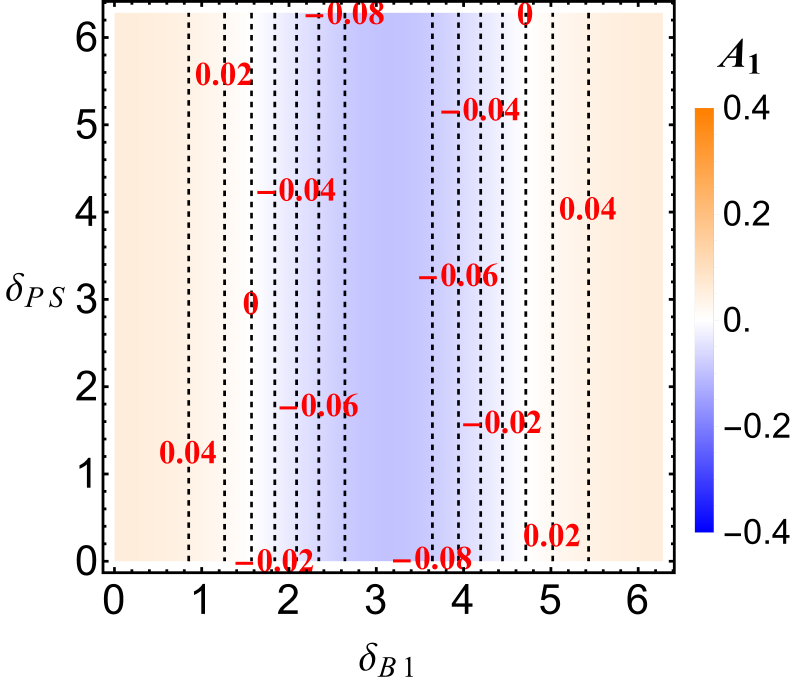}
    \includegraphics[keepaspectratio,width=5.4cm]{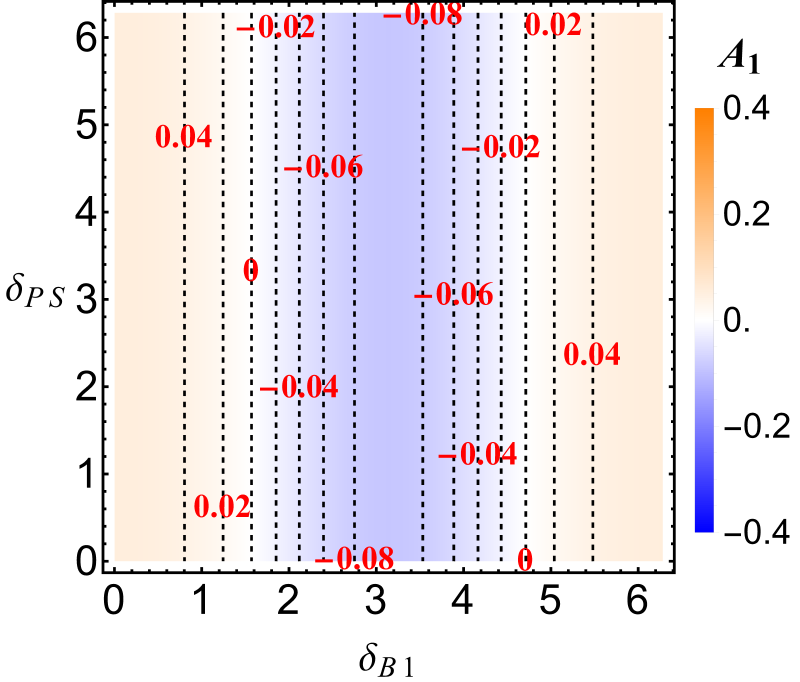}
    \caption{Two-dimensional dependence of the asymmetry $A_1$ on the strong phases in case 3. The left, middle, and right panels correspond to the three benchmarks (a) $r_{B1}=r_{B3}= r_0/2, r_{B2}=r_{B4}=2 r_0$, (b) $r_{B1}=r_{B2}=r_{B3}=r_{B4}=r_0$, and (c) $r_{B1}=r_{B3}=2 r_0, r_{B2}=r_{B4}= r_0/2$. The dashed lines on the graph represent contours with associated red values of $A_1$.}\label{fig:Case3A1}
\end{figure}

\begin{figure}
    \centering
    \includegraphics[keepaspectratio,width=5.4cm]{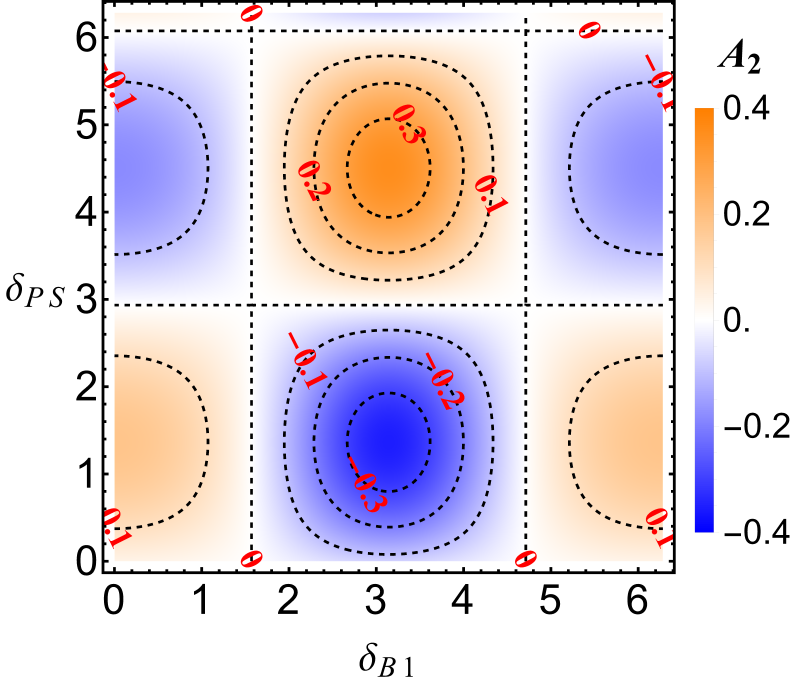}
    \includegraphics[keepaspectratio,width=5.4cm]{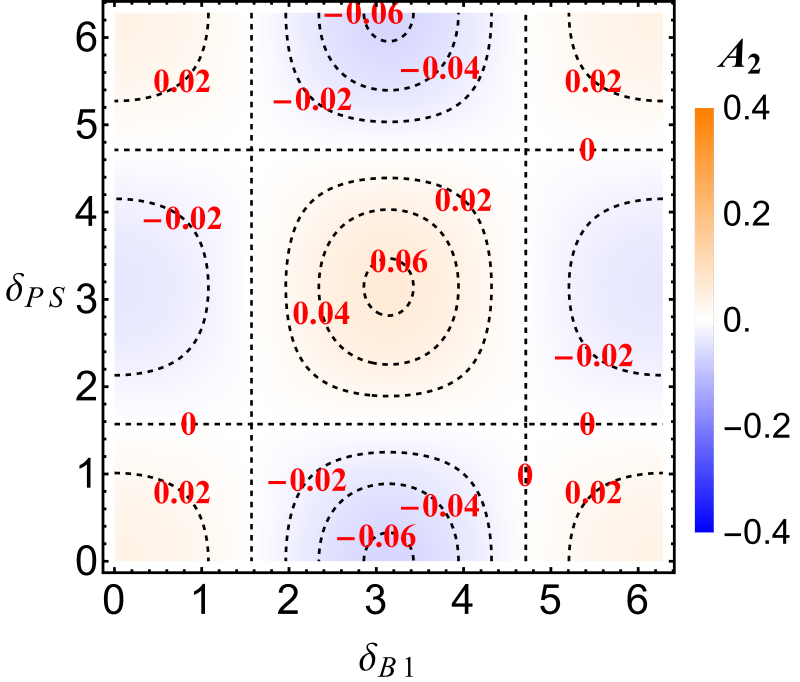}
    \includegraphics[keepaspectratio,width=5.4cm]{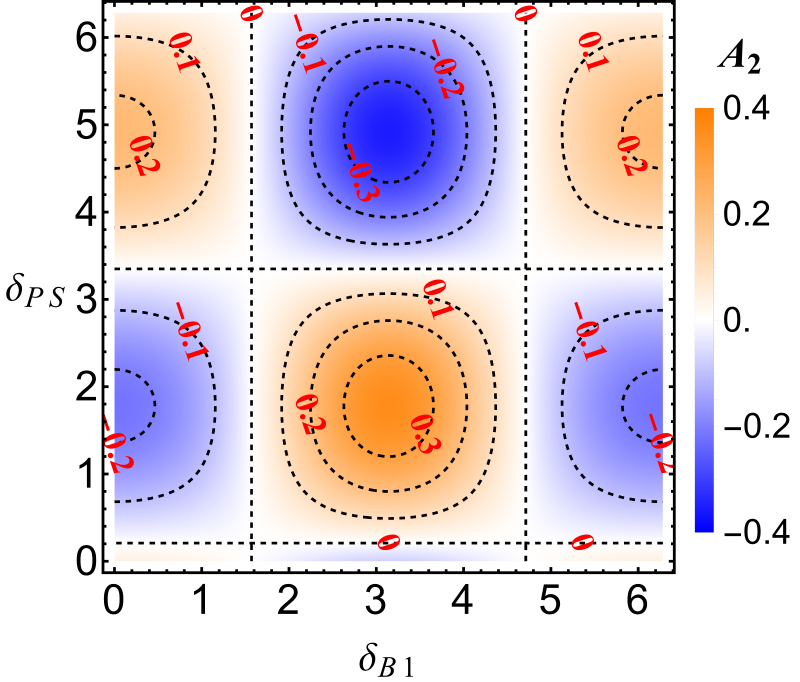}
    \caption{Two-dimensional dependence of the asymmetry $A_2$ on the strong phases in case 3. The left, middle, and right panels correspond to the three benchmarks (a) $r_{B1}=r_{B3}= r_0/2, r_{B2}=r_{B4}=2 r_0$, (b) $r_{B1}=r_{B2}=r_{B3}=r_{B4}=r_0$, and (c) $r_{B1}=r_{B3}=2 r_0, r_{B2}=r_{B4}= r_0/2$. The dashed lines on the graph represent contours with associated red values of $A_2$.}\label{fig:Case3A2}
\end{figure}

{\bf Case 3.}
Finally, in case 3, we explore a scenario identical to case 2 except for setting the strong phases as $\delta_{B1}=\delta_{B2}=-\delta_{B3}=-\delta_{B4}$, and $\delta_{PS}=\delta_{DP}$. The dependencies of $A_{1,2}$ on the two strong phases are displayed in FIG.~\ref{fig:Case3A1} and FIG.~\ref{fig:Case3A2}, respectively. 

In this case, the contributions to $A_1$ from two different partial waves of resonances largely cancel each other out, resulting in relatively small typical values for $A_1$. This emphasizes the significance of $A_2$ as a complementary observable. With the exception of the second benchmark for the amplitude ratios, the $A_2$ values are generally quite substantial across the majority of the parameter space defined by the strong phases.

\end{widetext}

Despite the various patterns depicted in FIG.~\ref{fig:Case1} - FIG.~\ref{fig:Case3A2}, it is evident that the CP asymmetry can easily reach magnitudes of $\mathcal{O}(10\%)$ across a wide range of strong phases in most scenarios. Furthermore, in specific regions, the CP asymmetry can even exceed $50\%$. In cases where the direct CP asymmetry $A_1$ is small, the complementary information provided by the sizable $A_2$ offers additional insight into CP violation. Taking into account the relevant measurements from LHCb~\cite{LHCb:2013jih, LHCb:2021ohr} and the decay width of this process [as indicated in \eqref{Den}], we estimate that approximately $\mathcal{O}(10^2)$ events will be collected after the completion of Run 1-2 at LHCb, with even more expected in Run 3-4. Therefore, this channel presents a significant opportunity to uncover CP violations in the baryon sector. 

In practice, our analysis does not encompass the entire contributions of both $N(1440)$ and $N(1520)$ due to their wide decay widths. Instead, our study applies to the phase space with a specific value of the $p\pi^-$ invariant mass and the neighbor region. Interestingly, this situation presents an advantage as different $p\pi^-$  invariant masses correspond to distinct strong phases, and it is highly probable that in certain regions these strong phases would result in substantial CP asymmetries.

At the end of this section, it is noteworthy that there are also other decay channels potentially exhibiting significant magnitudes of CP violation induced by a similar mechanism. One example is $\Lambda_b \to D(\to K^+ K^-) \Lambda$, whose amplitudes with intermediate $D^0$ and $\bar{D}^0$ also have very large weak phase differences. When considering the magnitudes of the relevant CKM matrix elements and their corresponding decay branching ratios~\cite{ParticleDataGroup:2022pth, LHCb:2013jih, LHCb:2021ohr}, we estimate that the branching ratio can be several times bigger than that of $\Lambda_b \to D(\to K^+\pi^-) N(\to p\pi^-)$. Therefore, we recommend that $\Lambda_b \to D(\to K^+K^-) \Lambda$ is another candidate channel for baryonic CP-violation discovery.

{\it Conclusion.---} 
We have systematically investigated the CP violation in the $\Lambda_b \to D(\to K^+\pi^-) N(\to p\pi^-)$ decay.  The large weak phase difference between the contributing amplitudes from $D^0$ and $\bar{D}^0$ suggests the potential for significant CP violation in this process. While the CP asymmetry is influenced by various parameters, including strong phases and amplitude ratios, our phenomenological analysis demonstrates that CP violation remains substantial across a wide range of parameter values. Considering the luminosity and reconstruction efficiency of the LHCb experiment, we estimate that the reconstruction of approximately $\mathcal{O}(10^2)$ events is feasible during the Run 1-2 phase, with even more events expected in Run 3-4. Based on these findings, we strongly recommend that the LHCb Collaboration prioritizes the investigation of baryonic CP violation using $\Lambda_b \to D(\to K^+\pi^-) N(\to p\pi^-)$ and similar channels such as $\Lambda_b \to D(\to K^+ K^-) \Lambda$.

{\it Acknowledgements}
The authors are grateful to Bo Fang, Wen-Bin Qian, Liang Sun, Yue-Hong Xie, Fu-Sheng Yu, and Zhen-Hua Zhang for their insightful discussions pertaining to both the experimental and theoretical aspects of this study. 
This work is supported in part by the National Natural Science Foundation of China under Grants No. 12005068, 12375086, 11975112, and 12335003; National Key Research and Development Program of China under Contract No. 2020YFA0406400; and the Fundamental Research Funds for the Central Universities under Grant No.lzujbky-2023-it12.

\end{document}